\documentclass[11pt,a4paper]{article}                           

\usepackage{amsmath}
\usepackage{amssymb}
\usepackage[usenames]{color} 
\usepackage{multirow}
\usepackage{graphicx}
\usepackage{epstopdf}
\usepackage{array}
\usepackage{hhline}
\usepackage{cite}
\usepackage{microtype}
\usepackage[bf]{caption}
\usepackage{color}
\usepackage[usenames,dvipsnames]{xcolor}
\usepackage{subcaption}
\usepackage{pstool}

\usepackage{ifpdf}
\ifpdf
  \usepackage[pdftex,
    pdftitle={},
    pdfauthor={},
    pdfsubject={},
    bookmarksopen, bookmarksnumbered, bookmarksopenlevel=2]{hyperref}
\fi
\def\hybrid{\topmargin -20pt    \oddsidemargin 0pt
        \headheight 0pt \headsep 0pt
        \textwidth 6.25in       
        \textheight 9 in       
        \marginparwidth .875in
        \parskip 5pt plus 1pt 
          \jot = 1.5ex
   }
\hybrid
\numberwithin{equation}{section}
\numberwithin{table}{section}

\newcommand{\beq}{\begin{equation}}  \newcommand{\eeq}{\end{equation}}
\newcommand{\bal}{\begin{aligned}}   \newcommand{\eal}{\end{aligned}}
\newcommand{\bea}{\begin{eqnarray}}  \newcommand{\eea}{\end{eqnarray}}

\newcommand{\bmat}{\left(\begin{array}}
\newcommand{\emat}{\end{array}\right)}

\newcommand{\nn}{\nonumber}

\usepackage{cancel}

\newcommand{\be}{\begin{equation}}
\newcommand{\ee}{\end{equation}}

\newcommand{\half}{\frac{1}{2}}

\begin{document}

\baselineskip=14pt
\parskip 5pt plus 1pt

\vspace*{-1.5cm}
\begin{flushright}    
  {\small 
  LMU-ASC 55/17 \\
  MPP-2017-185
  }
\end{flushright}

\vspace{2cm}
\begin{center}        
  {\LARGE Scalar Fields, Hierarchical UV/IR Mixing
  \\ 
  \vspace{0.2cm}
  and The Weak Gravity Conjecture}
\end{center}

\vspace{0.5cm}
\begin{center}        
{\large Dieter L\"ust$^{1,2}$ and Eran Palti$^2$}
\end{center}

\vspace{0.15cm}
\centerline{$^1$\it  Arnold--Sommerfeld--Center for Theoretical Physics,}
\centerline{\it Ludwig--Maximilians--Universit\"at, 80333 M\"unchen, Germany}
\vspace{0.15cm}
\begin{center}        
  \emph{$^2$Max-Planck-Institut f\"ur Physik (Werner-Heisenberg-Institut), \\
Fohringer Ring 6, 80805 Munchen, Germany}
             \\[0.15cm]
 
\end{center}

\vspace{2cm}


\begin{abstract}
\noindent
The Weak Gravity Conjecture (WGC) bounds the mass of a particle by its charge. It is expected that this bound can not be below the ultraviolet cut-off scale of the effective theory. Recently, an extension of the WGC was proposed in the presence of scalar fields. We show that this more general version can bound the mass of a particle to be arbitrarily far below the ultraviolet cut-off of the effective theory. It therefore manifests a form of hierarchical UV/IR mixing. This has possible implications for naturalness. We also present new evidence for the proposed contribution of scalar fields to the WGC by showing that it matches the results of dimensional reduction. In such a setup the UV/IR mixing is tied to the interaction between the WGC and non-local gauge operators. 
\end{abstract}

\thispagestyle{empty}
\clearpage

\setcounter{page}{1}


\newpage

\tableofcontents

\section{Introduction}

The gauge hierarchy problem can be phrased as the question of how ultraviolet (UV) physics can lead to a scalar field mass which is far below the UV scale. Separating an infrared (IR) scalar mass from the UV scale requires a very intricate cancellation between all the contributions to the mass. The standard proposed solutions to this problem are typically based on introducing new physics near the scalar mass. This can either be a new symmetry, like supersymmetry, or more drastically Quantum Gravity physics, for example large extra dimensions \cite{ArkaniHamed:1998rs}, or the more general formulation in terms of a large number of species \cite{Dvali:2007hz}. However, the absence of any experimental signs of new physics at energy scales near the mass of the Higgs motivates thinking about the possibility that there could be some property of UV physics which is responsible for the cancellation between the contributions to the scalar mass even when the UV scale is far away. Such a property would have to belong to UV physics and yet manifest as a restriction on the mass of a scalar at a much lower IR scale. In this sense it should exhibit some form of UV/IR mixing. 

It is expected that quantum gravity manifests UV/IR mixing. The classic example being that increasingly massive black holes have increasingly large horizon areas. One conjectured property of quantum gravity is the Weak Gravity Conjecture (WGC) \cite{ArkaniHamed:2006dz}. It is therefore interesting to ask if the WGC could manifest some sort of UV/IR mixing such that it restricts the mass of a scalar to an IR scale far below the UV scale associated to quantum gravity physics. In the original formulation of \cite{ArkaniHamed:2006dz} this was not possible. In this note we present evidence that for the more general formulation in the presence of scalar fields \cite{Palti:2017elp} it is, at least in principle, possible. Specifically, we propose that quantum gravity physics can restrict the mass of a particle to an IR scale far below the UV cut-off scale of an effective theory if the particle couples to gauge fields and massless scalar fields with almost precisely equal strength. 

The rest of the introduction is dedicated to expanding on this and forms an overview of some of the main points of this note. Consider a theory with gravity, a $U(1)$ with gauge coupling $g$, a charged scalar $h$ with charge $q=1$ and mass $m$, and a neutral scalar $\phi$ with mass $m_{\phi}$
\be
{\cal L} = \frac{M_p^2}{2} R -\frac{1}{4 g^2} F^2 - \left|D h \right|^2 - \half \left(\partial \phi \right)^2 - m^2 h^* h - \half m_{\phi}^2 \phi^2 - 2 m \mu \phi h^* h + ...  \;. \label{toymo}
\ee
Here $\mu$ is a dimensionless parameter which parameterises the coupling of $\phi$ to $h$. The $...$ denote arbitrary further terms in the Lagrangian. All the quantities in the Lagrangian (\ref{toymo}) are the quantum corrected expressions having integrated out all the UV physics. The UV completion of the theory is taken to be as follows. There is a scale $\Lambda_{UV} \sim g M_p$ above which the theory can not be completed by a quantum field theory, so it is the scale where quantum gravity related physics is reached. Below this scale, but above $m$ and $m_{\phi}$, there may be other scales where new degrees of freedom appear but for each one the theory can be completed by a quantum field theory. The new degrees of freedom are constrained such that the particle with the largest charge-to-mass ratio with respect to the $U(1)$ remains $h$. Then the claim is that in the limit $m_{\phi} \rightarrow 0$, this theory must satisfy \cite{Palti:2017elp}
\be
 \sqrt{g^2 - \mu^2}  M_p \equiv \beta M_p \geq m \;. \label{exmod}
\ee
We also impose $g \geq \mu$. For finite $m_{\phi} \neq 0$ the expression (\ref{exmod}) will receive corrections suppressed by $\frac{m_{\phi}}{m}$. We estimate these to be of order $\frac{\mu^2 m_{\phi}}{\beta m}M_p$. We therefore observe a new mass scale in the theory $\beta M_p$. We will argue that, assuming sufficiently small $m_{\phi}$, this mass scale can be separated arbitrarily far from the quantum gravity UV scale of the theory $m \leq \beta M_p \ll \Lambda_{UV}$. Since (\ref{exmod}) is tied to quantum gravity physics, with an associated mass scale $\Lambda_{UV}$, but is a constraint on an arbitrarily low IR scale, it manifests a form of UV/IR mixing.

In other words, say the particle with the largest charge-to-mass ratio in the theory happens to couple almost precisely the same to gauge fields and massless scalar fields. From the perspective of quantum field theory this would not have any implications, but from the perspective of quantum gravity we claim that this would imply that the mass of the particle would have to be at an IR scale far below the UV scale of the theory. The reason is that if $m$ violates the bound (\ref{exmod}) then the charged particle $h$ would form a tower of absolutely stable bound states of increasing charges. While this will not lead to a direct violation of any physical principle, the bound states are similar to black hole remnants and would go against some expectations of quantum gravity.

There are two important and related points to state from the outset. The first is that what we propose is UV/IR mixing is different from other known cases. Specifically, one definition of UV/IR mixing is that it is a property of UV physics that naturally and inevitably one recovers an IR mass scale from it. For example, T-duality in string theory implies that going to the UV automatically recovers IR physics. This is not the case here. Rather, we have to input an IR mass scale $m_{\phi}$ into the theory, and further tune $\beta \ll g$. Nonetheless, (\ref{exmod}) could only be deduced, to our knowledge, from quantum gravity physics. This has an associated mass scale of the UV scale $\Lambda_{UV}$ which as $\frac{\beta}{g} \rightarrow 0$ becomes arbitrarily far away from the IR scale where this relation holds. We therefore believe this should be called UV/IR mixing.

The second point, which is a consequence of the first, is that the UV/IR mixing does not form in itself a solution to the problem of the naturalness of a scalar mass. This is because we have to put in an IR mass scale $m_{\phi}$ by hand and tune $\beta$ to obtain an IR bound on the particle mass. A full solution to the naturalness problem from UV/IR mixing would be one where the UV physics automatically gives an IR scalar mass. In other words, we propose that {\it if} a particle couples almost equally to gauge and light scalar fields {\it then} its mass is light. There is nothing stopping the particle being heavy and coupling to gauge and scalar fields differently. Our results are therefore only a reformulation of the question of the naturalness of a scalar mass. Interestingly, this reformulation is one which we believe could only arise from quantum gravity and involves UV/IR mixing in the sense described above. It therefore may have a role to play in a possible full solution to the naturalness problem that involves UV/IR mixing in some way.\footnote{Indeed some interesting recent proposals that are potentially relevant for naturalness and which utilise the WGC \cite{Ibanez:2017kvh,Hamada:2017yji,Ibanez:2017oqr} assume that it must have some UV/IR mixing properties.}

Indeed, a striking property of (\ref{exmod}) is that it is a bound on a dimensionful mass parameter by dimensionless  parameters. This is certainly encouraging in terms of technical naturalness. Both $g$ and $\mu$ are technically natural. The parameter $\beta$ is not technically natural in the sense of there being an enhanced symmetry when it vanishes, however, it is very well protected. Consider the theory (\ref{toymo}) but without any additional terms. Then the 1-loop correction to the gauge coupling diverges logarithmically but goes as $g^3$. The correction to $\mu$ is not divergent, and goes as $\mu^3$ and $g^2 \mu$. Then as long as $g^2 \sim \mu^2 \lesssim \beta$, the hierarchy $\beta \ll g$ is not disturbed. This tells us that if we recouple new physics then it does not need to disturb a hierarchy. Of course, it is easy to couple in new physics that does spoil a hierarchy $\beta \ll g$ by loop corrections, but the hierarchy is stable with respect to the physics needed for the bound (\ref{exmod}). 

We have also introduced a light neutral scalar $\phi$ by hand. While keeping it light after coupling to new physics beyond (\ref{toymo}) is generally difficult, within (\ref{toymo}) the scalar only has to couple to $h$, and so the contributions of loops of $h$ to its mass will go as $m_{\phi} \sim \mu m$. Since $\mu$ is small the scalar remains light with respect to $m$. Of course, if $\phi$ couples directly to some new high mass scale then it is expected to gain a large mass, but no such coupling is required for (\ref{exmod}). Rather, it is a model-building challenge to UV complete (\ref{toymo}) in a way that allows $\phi$ to couple much more strongly to $h$ than to high scale UV physics.

The note is structured as follows. In section \ref{sec:WGCUVIR} we discuss generally UV/IR mixing and the Weak Gravity Conjecture. In section \ref{sec:gwgc} we develop a toy model based on dimensional reduction of a 5-dimensional theory. This serves the role of illustrating the relevant physics, but also leads to new insights on the WGC. In section \ref{sec:cutoff} we discuss the UV cut-off scale associated to quantum gravity physics. In section \ref{sec:QC} we discuss quantum corrections and the implications for naturalness. We finish with a discussion in section \ref{sec:summary}.

\section{The Weak Gravity Conjecture and UV/IR Mixing}
\label{sec:WGCUVIR}

The Weak Gravity Conjecture (WGC) \cite{ArkaniHamed:2006dz} (see \cite{Rudelius:2015xta,Brown:2015iha,Bachlechner:2015qja,Montero:2015ofa,Hebecker:2015rya,Brown:2015lia,Heidenreich:2015wga,Palti:2015xra,Heidenreich:2015nta,Kooner:2015rza,Ibanez:2015fcv,Hebecker:2015zss,Hebecker:2015tzo,Conlon:2016aea,Long:2016jvd,Heidenreich:2016aqi,Montero:2016tif,Hebecker:2016dsw,Saraswat:2016eaz,Herraez:2016dxn,Ooguri:2016pdq,Hebecker:2017uix,Cottrell:2016bty,Hebecker:2017wsu,Dolan:2017vmn,Ibanez:2017kvh,Cvetic:2017epq,Hamada:2017yji,Ibanez:2017oqr,Montero:2017mdq,toappear} for some recent work on this) states that given a $U(1)$ symmetry with gauge coupling $g$, there must exist a particle of charge $q$ and mass $m$ which satisfies the inequality
\be
g q M_p \geq m \;. \label{WGC}
\ee 
We will henceforth generally set $q=1$. This bound is interesting in the sense that even though it is based on quantum gravity physics, it can bound the mass of a particle to be parametrically light compared to the Planck scale. It is therefore natural to consider if the bound on the WGC particle mass could be manifesting UV/IR mixing. One property of the WGC that could be in support of this is that it appears to be in tension with Wilsonian naturalness. The left hand side of (\ref{WGC}) runs logarithmically with energy scales, but if the WGC particle is a scalar, then it is bounding a mass that runs linearly with the energy scale. This means that the scalar mass $m$ is bounded to be only logarithmically sensitive to UV physics at energy scales above $g M_p$. 

However, it is expected that (\ref{WGC}) does not manifest UV/IR mixing, and for the same reason is also consistent with the Wilsonian notion of naturalness \cite{Cheung:2014vva}. The reason is that it is expected that the cut-off scale of the effective theory $\Lambda$ is bounded by 
\be
g M_p \geq \Lambda \;. \label{WGCcutoff}
\ee
Indeed, (\ref{WGCcutoff}) can be argued for by reasoning unrelated to naturalness and is termed the Magnetic WGC \cite{ArkaniHamed:2006dz}. Therefore the WGC (\ref{WGC}) only bounds the mass to be below the UV cut-off scale of the effective theory. While it is very interesting that the cut-off scale $\Lambda$ can be far below the expected quantum gravity physics scale $M_p$, within the effective theory $g M_p$ is the UV scale.

Recently a generalisation of the WGC in the presence of massless scalar fields was proposed \cite{Palti:2017elp}. Restricted to a simplified case so as to be comparable with (\ref{WGC}) it states that 
\be
\left(g^2 - \mu^2 \right)M^2_p  \geq m^2 \;. \label{GWGC}
\ee
Here $\mu$ is the coupling of the WGC particle to massless scalar fields. For example, if we consider a single real canonically normalised massless scalar $\phi$, and expand it about its vacuum expectation value $\phi=\left<\phi\right>+\delta \phi$, then the coupling to a complex scalar WGC particle $h$ arises from the Lagrangian term
\be
{\cal L} \supset 2 m \mu \delta \phi h^* h \;. 
\ee
Similarly, for a fermion WGC particle $\psi$ the coupling is
\be
{\cal L} \supset \mu \delta \phi \overline{\psi} \psi \;.
\ee
Note that this can also be written as $\mu =\partial_{\phi} m$, which is how it was presented in \cite{Palti:2017elp}. 

The key point of interest for this note is that in the presence of massless scalar fields it is possible to bound the mass $m$ to an IR mass scale by sending 
\be
\beta^2 \equiv g^2 - \mu^2 \rightarrow 0 \;, \label{GWGCLimit}
\ee
while keeping $g$, and therefore $\mu$, finite. If the UV scale associated to quantum gravity physics is still only constrained by (\ref{WGCcutoff}), then for $\beta \ll g$ the WGC particle mass is bound far below the UV scale of the effective theory $m \ll \Lambda$. Since the WGC is associated to UV quantum gravity physics but is bounding a mass at an IR scale, this would be a manifestation of UV/IR mixing. 

There are two crucial questions with regards to this proposal. The first is whether (\ref{GWGC}) is indeed a property of quantum gravity. The second is whether sending $\beta \rightarrow 0$ does not also modify the cut-off scale of the effective theory so that the quantum gravity UV scale would be $\beta M_p$ rather than $g M_p$. If that happened then we would be back to the case where the bound on the mass is only the UV cut-off scale of the effective theory. In this note we present new evidence that (\ref{GWGC}) is a property of quantum gravity and that sending $\beta \rightarrow 0$ does not lower the quantum gravity cut-off of the effective theory. Note that it is possible that $\beta \rightarrow 0$ does lower the scale of some physics, like a finite number of particles, but we claim that the scale at which one must utilise quantum gravity physics can remain finite.  

In \cite{Palti:2017elp} the bound (\ref{GWGC}) was proposed based on a number of reasons. One is related to ${\cal N}=2$ supersymmetry and is discussed below. A more general argument is that if we consider the particle with the largest charge to mass ratio in the theory, then unless it satisfies (\ref{GWGC}) it would form a tower of stable gravitationally bound states. Specifically, two such particles would feel a mutual repulsive force due to the gauge field with strength $g$, an attractive force due to gravity with strength $m$, and a further attractive force due to the massless scalar fields with strength $\mu$. The inequality (\ref{GWGC}) is the requirement that the repulsive force beats the attractive forces so that the particles do not form a bound state. Such a bound state would be completely stable by charge and energy conservation. We could then keep building such stable states by adding more particles. 

Now consider how many such bound states we could fit below a finite mass scale, say $M_p$. We know that the particle mass must satisfy $m \geq \beta M_p$ to form bound states. Then we can fit at least $\frac{1}{\beta}$ such states below $M_p$. But now send $\beta \rightarrow 0$ and we find an infinite number of states. By a species argument, as in  \cite{Dvali:2007hz} for example, we deduce that the cut-off scale of the theory must go to zero in this limit. However, we have proposed, and will present evidence for this in section \ref{sec:cutoff}, that the cut-off scale of the theory stays finite in this limit. This therefore implies that the stable states must not be present to start off with, and so (\ref{GWGC}) must be satisfied. Note that one way to avoid this is to say that we are free to take $m \gg \beta M_p$. This is true, but nonetheless, we can imagine starting with $m < \beta M_p$ and adiabatically increasing $m$ past the threshold $\beta M_p$ turning a theory consistent with quantum gravity into a theory inconsistent with quantum gravity and therefore $\beta M_p$ is clearly an intrinsic quantum gravity scale. 

We can reach the same conclusion by thinking about black hole remnants \cite{Palti:2017elp}. If an extremal black hole sources scalar fields through its charge, then the relation between its charge and ADM mass is modified in such a way that, in principle, the ADM mass could be far smaller than the charge. The relation between the ADM mass and the charge is the same as (\ref{GWGC}), at least for any black holes that can be described by a so-called `fake superpotential' \cite{Ceresole:2007wx}, but with the scalar coupling $\mu$ replaced by an appropriate coupling to the black hole. Schematically, if the gauge coupling depends on some massless scalar field $g\left( \phi \right)$, and if we can write it as
\be
g\left( \phi \right)^2 = W\left(\phi\right)^2 + \left( \partial_{\phi} W \right)^2 \;, \label{fakes}
\ee
for some real function $W\left(\phi\right)$, then the black hole solution will be such that the ADM mass is $M_{ADM}=W\left(\phi_{\infty}\right)$. Here $\phi_{\infty}$ is the value of the field at spatial infinity. We can therefore define
\be
\beta_{BH}^2 \equiv g\left( \phi_{\infty} \right)^2 - \left. \left( \partial_{\phi} W \right)^2\right|_{\phi=\phi_{\infty}}\;,
\ee 
and write $M_{ADM} = \beta _{BH} M_p$. Now one can consider how many black hole remnants can fit below $M_p$ and again we find $\frac{1}{\beta _{BH}}$ such states. By sending $\beta_{BH} \rightarrow 0$ we can reach an infinite number of remnants but with a finite cut-off. This is inconsistent and therefore the remnants must be able to decay. 

While this argument tells us that there should be a particle in the theory that must become arbitrarily super-extremal in order for black holes to decay and avoid the remnants, it does not quite imply (\ref{GWGC}) since we need some relation between $\beta_{BH}$ and $\beta$ for this. There is a nice way to see such a relation, and also to connect the remnant picture with the bound states picture. It is possible to think of the extremality bound on a black hole as the statement that the black hole mass should be at least the binding energy in bringing in all the charge in the black hole from infinity \cite{Banks:2006mm}. In the presence of scalar fields this binding energy decreases since the attractive force makes it easier to bring the charges in. This is why the ADM mass of a black hole can be much less than its charge. It also directly relates the bound states with the remnants, and justifies a relation between $\beta_{BH}$ and $\beta$. While it is important to find a more quantitative version of this relation, we believe that (\ref{GWGC}) is strongly motivated by this argument.\footnote{Quantitatively, we are assuming that $q_{\mathrm{particle}} \beta_{BH} \leq q_{BH} \beta$, where $q$ denotes the charge. The more general expression is ${\cal Q}^2_{\mathrm{particle}} \beta^2_{BH} \leq {\cal Q}^2_{BH} \beta^2$, with ${\cal Q}^2$ defines as in \cite{Palti:2017elp}. }

Note that the remnants argument is in some sense stronger than the bound states one, but requires the additional assumptions about the dependence of $g$ (and thereby $\mu$) on $\phi$. We can therefore define two versions of the general WGC (\ref{GWGC}). The first one is assuming the structure of $g\left( \phi \right)$ (\ref{fakes}) and is therefore motivated by both remnants and bound states. The second one is a stronger statement which does not assume anything about the dependence of $g$ on $\phi$, and is only motivated by bound states. We denote the former the Weak General WGC and the latter the Strong General WGC. Note that if one adopts only the weak version, then the action (\ref{toymo}) should be appropriately modified. This can lead to an additional source of mass for the field $\phi$ and should be accounted for in understanding $m_{\phi}$ within a UV completion.  

It is worth noting that these arguments are much stronger than those presented in \cite{ArkaniHamed:2006dz} for the original WGC. There the number of bound states, or remnants, below $M_p$ was $\frac{1}{g}$. This is consistent with a species argument as long as the cut-off of the theory went to zero when $g \rightarrow 0$, which is indeed the case. Therefore the presence of the bound states or remnants was not excluded by any known arguments.

At this point it is worth making a further clarifying remark. In our analysis we will assume that the WGC particle has charge of order one $q \sim 1$. This is a strong version of the WGC, and it avoids possible loopholes to do with increasing the charge and mass of the particle so that it ends up above the cut-off scale of the effective theory. Such a possibility for avoiding a strong statement by the WGC was pointed out already in \cite{ArkaniHamed:2006dz}. It was also noted within the context of axion alignment \cite{Rudelius:2015xta,Brown:2015iha,Hebecker:2015rya}, and its dual version of higgsing a linear combination of $U(1)$ groups \cite{Saraswat:2016eaz}. This is supported by evidence from string theory \cite{Palti:2015xra, Heidenreich:2015nta,Heidenreich:2016aqi,Montero:2016tif}. The arguments presented above also rule out this possibility in the sense that if the WGC particle is kept at the cut-off scale of the theory, then we would still have an infinite number of bound states or remnants in the $\beta \rightarrow 0$ limit. 

\subsection{Relation to ${\cal N}=2$ Supersymmetry}

The general version of the WGC (\ref{GWGC}) has close ties to ${\cal N}=2$ supersymmetry. This is because BPS states in ${\cal N}=2$ saturate the bound (\ref{GWGC}). Indeed, the simplest way to argue for (\ref{GWGC}) is that in ${\cal N}=2$ supergravity supersymmetric black holes are themselves BPS states. This means that they can only decay to other BPS states. So the WGC particle, responsible for the decay of extremal black holes, must be a BPS particle and therefore satisfies (\ref{GWGC}). 

The ${\cal N}=2$ setting is also the simplest illustration of the UV/IR mixing property of (\ref{GWGC}). Consider going to a point in moduli space where the mass of a BPS state is vanishingly small $m \rightarrow 0$ but where $g$ remains finite. Since a BPS state saturates (\ref{GWGC}) this is also a point in moduli space where $\beta \rightarrow 0$ for that state. Now, consider just a pure gauge theory with no charged matter at this point in moduli space. This would be perfectly fine from the perspective of quantum field theory. However, the WGC would demand the existence of a charged particle. The original version (\ref{WGC}) would tell us that this charged particle must have a mass below $g M_p$. However, we know that the decay of black holes requires a much stronger condition which is that the charged particle must actually be BPS. This implies that its mass is at an IR scale far below $g M_p$ and therefore, as we argue in section \ref{sec:cutoff}, far below the scale of quantum gravity physics. The general WGC (\ref{GWGC}) is not quite as strong a statement as ${\cal N}=2$ would imply, since it could be satisfied in principle by a non-BPS particle, but the UV/IR mixing aspect is the same.

There is a also a more practical relation between (\ref{GWGC}) and ${\cal N}=2$ supersymmetry in that it is a setting that allows for protection against quantum corrections. Of course with respect to the hierarchy problem this is not so interesting since supersymmetry also protects a scalar mass. But from the perspective of trying to understand the microscopic physics behind (\ref{GWGC}) it is a useful starting point.

\section{The WGC with Scalars and Dimensional Reduction}
\label{sec:gwgc} 

In this section we present new evidence for (\ref{GWGC}) based on dimensional reduction. The point is that the WGC (\ref{WGC}) in 5 dimensions leads, upon a classical dimensional reduction, to the generalised version of the WGC (\ref{GWGC}) in 4 dimensions.\footnote{See also \cite{Heidenreich:2015nta,toappear} for a dimensional reduction analysis of the WGC.} In this section we will consider a classical dimensional reduction. In general, as discussed in the introduction, at the quantum level it is difficult to control $\beta$ and the mass of the scalar mediators. This example model is no exception to this, and quantum corrections can significantly modify the scenario. However, we will be able to control them, at least to some extent, by utilising some supersymmetry and will argue that in that case the key physics insights of the classical results will remain. We return to a discussion of this as part of the general discussion on quantum corrections in section \ref{sec:QCtm}. 

We consider first the case where the WGC particle is a scalar and return to the fermion case later. To simplify notation we henceforth work in units where $M_p=1$. The 5-dimensional theory of interest is 
\be
S_{5D} = \int_{M_4 \times S^1} d^5 X \sqrt{-G} \left[ \frac12 R^{(5)}  - \frac{1}{4g_5^2} F_{MN} F^{MN} - D_M H \left(D^M H \right)^* - M_H^2 H H^* \right] \;. \label{5daction}
\ee
Here, we have a higher dimensional gauge field $A^M$, with gauge coupling $g_5$, and a complex scalar $H$. The scalar $H$ is charged with charge $q$ and acts as the WGC particle to make this theory consistent. The bound on its mass from the 5-dimensional WGC reads (see for example \cite{Heidenreich:2015nta})
\be
g_5 q \geq \sqrt{\frac23}  M_H \;. \label{5dWGC}
\ee
If we reduce this theory on a circle, then we get an effective 4-dimensional action for the zero modes 
\be
S^0_{4D} = \int_{M_4} d^4 x \sqrt{-g} \left[ \frac12 R + {\cal I}_{IJ}  F_{\mu\nu}^{I}  F^{J,\mu\nu}  - \frac12 \left(\partial \varphi \right)^2 - \frac{1}{2 r^2g_5^2} \left(\partial a \right)^2 - \left| D h \right|^2 - m^2 h^* h\right] \;. \label{4dreducedaction}
\ee
Here the 4-dimensional fields are the dilaton $\varphi$, an axion $a$, a complex scalar $h$, the graviphoton $A_0^{\mu}$, and the zero mode of the gauge field $A_1^{\mu}$. The dilaton measures the circumference $r$ of the extra dimension
\be
r = e^{-\sqrt{\frac{2}{3}}\varphi} \;.
\ee 
The covariant derivative of the scalar field is given by
\be
D_{\mu} h = \left(\partial_{\mu} - i q A^1_{\mu} \right) h \;,
\ee
where $q$ is its charge, and the mass is given by
\be
m^2 =  \frac{M_H^2}{r} +\frac{q^2 a^2}{r^3} \;.
\ee
The index $I=0,1$ runs over the two 4-dimensional gauge fields $A_I^{\mu}$ and reads
\be
{\cal I}_{IJ} = -\frac{r}{4 g_5^2} \left( \begin{array}{cc} \frac{r^2g_5^2}{2} + a^2 & -a \\ -a & 1 \end{array} \right) \;,\;\; \left({\cal I}^{-1}\right)^{IJ} = -\frac{8}{r^3} \left( \begin{array}{cc}  1 & a \\ a & \frac{r^2g_5^2}{2} + a^2 \end{array} \right)\;. 
\ee
The axion $a$ originates from the extra-dimensional component of the gauge field $a=\int_{S^1}A_4$. It can be thought of as a Wilson line in the extra dimension and it enjoys a gauge discrete shift symmetry. There are no fields charged under the graviphoton $U(1)_0$ since we only write the action for the zero modes. The WGC states for $U(1)_0$ are KK modes. The lightest field charged under $U(1)_1$ is the scalar $h$ and so it plays the role of the WGC particle. 

To utilise the WGC (\ref{GWGC}) we need the general formulation for it presented in \cite{Palti:2017elp}. For a Lagrangian (not including the WGC particle $h$)  
\be
 \frac{R}{2} - g_{ij}\partial_{\mu} \phi^i \partial^{\mu} \phi^{j} + {\cal I}_{IJ} F_{\mu\nu}^{I}  F^{J,\mu\nu} +  {\cal R}_{IJ} F_{\mu\nu}^{I} \left(\star  F\right)^{J,\mu\nu} \;,    \label{genaction}
\ee
with scalar fields $\phi^i$ and gauge fields $A^I$, the generalisation of  (\ref{GWGC}) is
\be
{\cal Q}^2 \geq m^2 + g^{ij}\partial_{i} m \partial_{j} m \;. \label{gnsWGC}
\ee
For purely electric charges $q_I$, we have ${\cal Q}^2 = -\frac12 q_I \left({\cal I}^{-1}\right)^{IJ} q_J$. Applying the general expression (\ref{gnsWGC}) to (\ref{4dreducedaction}) we have
\be
{\cal Q}^2 \geq m^2 + \mu_a^2  + \mu_{\varphi}^2 \;, \label{gWGCmu}
\ee
where we define $\mu_a^2 = g^{aa} \partial_a m \partial_a m$, and $\mu_{\varphi}^2  = g^{\varphi\varphi} \partial_{\varphi} m \partial_{\varphi} m$. These measure the strength of the force mediated by the axion and dilaton respectively.
The full expressions are
\bea
{\cal Q}^2 &=& \frac{2 q^2 g_5^2}{r} + \frac{4a^2q^2}{r^3}\;,\;\;
m^2 = \frac{M_H^2}{r} + \frac{q^2 a^2}{r^3} \;,\nn \\
\mu_a^2 &=& \frac{2 q^4 g_5^2 a^2 }{r^4m^2} \;,\;\;
\mu_{\varphi}^2 = \frac{\left(\sqrt{\frac{2}{3}} \frac{M_H^2}{r} + \sqrt{6} \frac{q^2 a^2}{r^3} \right)^2}{2m^2} \;. \label{eqst}
\eea
Using these, the 4-dimensional bound (\ref{gWGCmu}) reads
\be
\frac{2M_H^2 r}{3\left(a^2 q^2 + M_H^2 r^2 \right)}\left(3 g_5^2 q^2 - 2 M_H^2  \right) \geq 0 \;. \label{4dbo}
\ee
This precisely reproduces the 5-dimensional bound (\ref{5dWGC}).

We therefore find that the general WGC with scalar fields can be deduced from a classical dimensional reduction of the WGC in the absence of scalar fields. This lends some further weight to its validity. It is also interesting in the sense that it forms an example where the bound (\ref{gnsWGC}) is not saturated.

For the case where the WGC particle is a fermion we can consider a 5-dimensional Dirac fermion $\Psi$, with 5-dimensional mass $M_{\Psi}$ and 4-dimensional Dirac zero-mode $\psi$. Dimensional reduction leads to exactly the same results as (\ref{eqst}), with $M_H \rightarrow M_{\Psi}$, and therefore again matches the 5-dimensional WGC. Since, in this respect, it is no different from the scalar case we do not reproduce the calculations here. It is, however, worth commenting about a point regarding the pseudo-scalar nature of the axion $a$. In both the scalar and fermion cases the expression for $\mu_a$, which is the non-relativistic force mediated by the axion, vanishes if $a$ has a vanishing expectation value (in which case the non-relativistic limit requires $M_{\Psi} ,M_H\neq 0$). In the scalar case this is readily seen from the action (\ref{4dreducedaction}) since there is no cubic coupling for $a=0$, as is guaranteed by parity conservation. In the fermion case there is a cubic coupling in the action $r^{-\frac32} qa \overline{\psi} \gamma^5 \psi$. However, the straight exchange of the axion using this coupling does not lead to a long range force in the non-relativistic limit. Indeed, for $a=0$, the leading force is a dipole force which is spin-dependent and scales as $l^{-4}$, with $l$ the separation distance (see for example \cite{Ferrer:1998ue}). The expression for $\mu_a$ is capturing an exchange of the axion with an insertion of its vacuum expectation value at each external leg. As in the case of a scalar, it is a long range non-relativistic force that is only present due to the spontaneous breaking of parity. 

\section{The UV Cut-off Scale}
\label{sec:cutoff}

The results of \cite{Palti:2017elp}, and those of section \ref{sec:gwgc}, lend support to the proposal that (\ref{GWGC}) is a property of quantum gravity. The next question with regards to whether (\ref{GWGC}) can bound $m$ far below the UV cut-off scale is whether it is possible to send $\beta \rightarrow 0$ without also implying $\Lambda \rightarrow 0$, where $\Lambda$ is the cut-off scale of the effective theory. In this section we present arguments that indeed this is possible.

Note that we take $\Lambda$ as an enforced cut-off scale on {\it any} quantum field theory due to the appearance of quantum gravity physics. Typically this would be the mass scale of an infinite number of states. So we consider $\Lambda$ as, for example, the string scale. It could be that for a given quantum field theory there is a lower cut-off scale than $\Lambda$ where new physics appears $\Lambda_{NP}$. But this lower cut-off could in principle be completed by some other quantum field theory. We do not make the claim that $\Lambda_{NP}$ does not go to zero when $\beta$ does. It could well be that there must be new physics at the scale $\beta M_p$ for a given quantum field theory. For example, supersymmetry may have to appear. But, although $\beta M_p$ is a scale implied by quantum gravity, we claim that this new physics is not that of quantum gravity.

Sending $g \rightarrow 0$ in (\ref{GWGC}) is clearly problematic in the sense that the gauge field decouples and we recover an exact $U(1)$ global symmetry. Note that taking multiple $U(1)$s, and therefore using the general expression (\ref{gnsWGC}) where $g^2 \rightarrow {\cal Q}^2$ does not help in this regard. Since in ${\cal Q}^2$ the charge vector $q$ is contracted with the gauge kinetic matrix ${\cal I}$, a vanishing ${\cal Q}^2$ would imply a vanishing determinant for ${\cal I}$ which would again decouple a gauge field leading to a global symmetry. This is expected to not be possible in quantum gravity. The bound (\ref{WGC}), or (\ref{GWGC}), alone would only imply that a particle become massless in this limit. This seems insufficient to block such a global symmetry, and it is therefore natural to expect that the stronger condition (\ref{WGCcutoff}) should be imposed. In contrast, the $\beta \rightarrow 0$ limit does not decouple the gauge field and therefore just a particle becoming massless appears consistent from this perspective.

In the presence of ${\cal N}=2$ supersymmetry the WGC particle must be BPS, at least in the sense that supersymmetric extremal black holes can only decay to BPS particles. BPS particles satisfy the equality limit of (\ref{GWGC}). Therefore in this case $m \rightarrow 0$ implies, rather than is just implied by, $\beta \rightarrow 0$. So if $\beta \rightarrow 0$ would imply $\Lambda \rightarrow 0$ then we would have a quantum gravity obstruction to massless BPS states. This would be very strange. Indeed, the classic example of BPS states in string theory are wrapped branes. If we consider compactifications of type IIB string theory on Calabi-Yau manifolds, then D3 branes wrapped on 3-cycles lead to BPS particles in four dimensions. If we go to the conifold locus in moduli space then one such BPS particle (hypermultiplet) becomes massless \cite{Strominger:1995cz}. But only one BPS state is becoming massless which can be understood by looking at its effect on the gauge coupling. While the conifold locus is a somewhat exotic point in complex-structure moduli space, there is no reason to expect that it is not possible to work with an effective theory with a finite cut-off in its vicinity. This is in contrast to the $g \rightarrow 0$ limit, where an infinite number of states become massless sending $\Lambda \rightarrow 0$ and matching general quantum gravity expectations. 

In section \ref{sec:gwgc} we showed that (\ref{GWGC}) can be understood from classical dimensional reduction. In such a setting there are a number of UV cut-off scales. There is the scale where the 5-dimensional theory breaks down (see for example \cite{Sundrum:2005jf}), the Kaluza-Klein scale, and an expected 5-dimensional magnetic WGC scale\footnote{Note that in these estimates we have included an extra factor of $r^{-1}$ relative to the versions of these expressions in the literature, due to the change from the 5-dimensional to 4-dimensional Planck mass. Either way, this extra factor is not important for the primary point.}
\be
\Lambda^2_{5D} \sim \frac{1}{r}\left(\frac{4\pi}{g_5}\right)^2 \;, \;\; \Lambda^2_{KK} \sim \frac{1}{r}\left(\frac{2\pi}{r}\right)^2 \;, \;\; \Lambda^2_{WGC,5} \sim \frac{g^2_5}{r}  \;. 
\ee
We would like to consider how the parameter $\beta^2$ in (\ref{GWGCLimit}), which bounds the mass $m^2$, compares to these scales. Again we will focus on the classical results, and return to quantum corrections in section \ref{sec:QCtm}. The explicit form is
\be
{\cal Q}^2 - \mu_a^2 -\mu_{\varphi}^2 \equiv \beta^2 = \frac{3 a^4 q^4+6 a^2 M_H^2 q^2 r^2+6 g_5^2 M_H^2 q^2 r^4-M_H^4 r^4}{3 r^3 \left(a^2 q^2+M_H^2 r^2\right)}\;.
\ee
There are two ways in which we could choose the parameters in this model to obtain a small $\beta$. One of them violates the WGC while the other does not. Therefore, the difference between them must arise due to interaction with UV physics and so is associated to UV/IR mixing. 

Consider first the case which violates the WGC. Let us set $a = 0$ so that we have
\be
\left.\beta^2\right|_{a=0} = \frac{6 g_5^2 q^2 - M_H^2}{3 r} \;.
\ee
Then if we forget about the 5-dimensional WGC (\ref{5dWGC}), we can take $M_H^2 \rightarrow 6 g_5^2 q^2$. In that case we have $\beta \rightarrow 0$. This limit is perfectly fine from a quantum field theory perspective and it leads to a large mass $m^2 \rightarrow \frac{6 g_5^2 q^2}{r}$. So from a quantum field theory perspective, there is nothing that stops us from considering a particle which couples equally to gauge and scalar fields, yet has a large mass. This exemplifies the expectation that the generality of the connection between the coupling equality $\beta=0$ and the mass of a particle is inherently quantum gravitational in nature.

Consider now the way to reach small $\beta$ while respecting the WGC. We take the parameter values
\be
r^2 g_5^2 q\gg a^2 q\gg g_5 M_H r^2 \;. \label{para}
\ee
We have that $\beta^2$ and ${\cal Q}^2$ read in this case 
\be
\beta^2 \simeq \frac{a^2 q^2}{r^3}\;,\;\; {\cal Q}^2 \simeq \frac{q^2 g^2_5}{r}\;.
\ee
Therefore, we find $\beta^2 \ll {\cal Q}^2$ in this regime. Since (\ref{para}) is compatible with the WGC, we indeed find that the mass of the WGC particle is bound far below any UV cut-off scale. This is contrast to the field theory setting discussed above. It is also in contrast to the scale ${\cal Q}^2$, which would bound the mass in the absence of scalar fields, and which can not be separated from $\Lambda^2_{WGC,5}$. We therefore conclude that the UV/IR mixing properties of the WGC impose that in a quantum theory of gravity having a small $\beta$ implies a small mass for the WGC particle. 

Let us return to the case $a=0$. If we do impose the 5-dimensional WGC (\ref{5dWGC}) then we have 
\be
\left.\beta^2\right|_{a=0} \geq \frac{3 g_5^2 q^2}{2r}.
\ee
This means that the WGC bound on the mass is tied to the UV scale and there is no UV/IR mixing. This is interesting from the microscopic perspective since it implies that the UV/IR mixing is tied to $\mu_a$ which in this case vanishes. The contribution $\mu_a$ due to the axion arises in the 5-dimensional theory as a Wilson line around the extra circle dimension. A Wilson line is a non-local gauge operator. Therefore we see that the UV/IR mixing aspect is manifesting due to the interaction of the WGC with non-local physics. This matches the general expectation that UV/IR mixing should be tied to non-local physics. If we have $a \neq 0$, then the UV/IR mixing is tied to the parameter range (\ref{para}). This can be written as $a q \gg M_H r$. If we were to take $M_H r$ very large the WGC particle would have a localised Compton wavelength and would not sense the compactness of the circle direction or the associated Wilson line. It would be sensitive to only local physics and accordingly we see that UV/IR mixing would not be possible.

Note that the WGC bound (\ref{GWGC}) can only be understood in terms of classical long range forces if the mass of the WGC particle is heavier than the mediator forces. But the $\beta \rightarrow 0$ limit is the one where the WGC state becomes massless and therefore must be approached with care. We should keep this in mind, but it does not form a fundamental obstruction for any realistic scenario where we are concerned with $\beta \ll g$ rather than $\beta =0$.

To summarise, in this section we presented evidence that the general formulation of the WGC (\ref{GWGC}) can exhibit a bound on the WGC particle mass parametrically far below the UV cut-off scale of the theory. In other words, that the WGC particle possesses the property that if it couples almost precisely the same to gauge fields and scalar fields then its mass is far below the UV scale of the theory. 

\subsection{Relation to a Scalar Weak Gravity Conjecture}
\label{sec:sWGC}

In \cite{Palti:2017elp} it was shown that in an ${\cal N}=2$ supersymmetric setting the scalar forces act stronger than gravity on the WGC particle. More precisely, for each scalar field there is one particle on which the scalar force acts stronger than gravity. This was proposed as a possible Scalar WGC. If we apply this to the zero mode $h$ it reads 
\be
\left(\mu_a^2 + \mu_{\varphi}^2\right) M_p^2 \geq m^2 \;. \label{swgc}
\ee
Utilising (\ref{eqst}) this implies
\be
 3 a^2 q^4 \left(a^2 + g_5^2 r^2 \right) \geq \left(M_Hr \right)^4  \;. \label{swgcv}
\ee
Therefore, for sufficiently large $M_H$ we find that the Scalar WGC is not satisfied by the WGC particle. So gravity is acting stronger than the scalar forces on the WGC particle. 

A possible interpretation of this result is that it is evidence against a Scalar WGC. If that is the case it could have interesting implications for large field excursions in quantum gravity. In \cite{Palti:2017elp} it shown that a scalar WGC could be a possible explanation for the exponentially decreasing mass of states for super-Planckian scalar field variations.\footnote{See \cite{Ooguri:2006in,Baume:2016psm,Klaewer:2016kiy,Valenzuela:2016yny,Blumenhagen:2017cxt,Hebecker:2017lxm} for recent work on this.} Here it appears that taking large $M_H$ could allow to violate this. As we will discuss in section \ref{sec:QC}, in a quantum setting the mass $M_H$ should be tied to supersymmetry breaking. In this sense it could be suggesting that looking at highly non-supersymmetric situations could allow for large field variations. 

Having stated this, it is the case still that even in this model the Scalar WGC is satisfied in the sense that there is still a particle on which gravity acts weaker than the scalar forces. This is because any Kaluza-Klein mode of the 5-dimensional field $H$ will satisfy (\ref{swgc}). This can be seen by noting that the replacement $q a \rightarrow q a + 2 \pi n$, where $n$ is an integer (the Kaluza-Klein number), is a gauge symmetry of the theory. For any non-zero $n$ the inequality (\ref{swgcv}) can not be violated within the effective theory limit $M_H r \ll 2\pi$.

Yet another possibility is to require that the Scalar WGC applied to the zero-mode $h$, which is the WGC particle in the sense that it has the largest charge-to-mass ratio, should be satisfied. Then this would be interpreted as a bound on the parameters (\ref{swgcv}). Or as a statement that the original 5-dimensional theory is not compatible with a quantum gravity UV completion as it is. 

With respect to the point of interest in this note. The Scalar WGC can also be used to restrict the mass of a particle since it imposes $\mu M_p \geq m$. This appears in many ways simpler than utilising (\ref{GWGC}). However, it is natural to expect that sending $\mu \rightarrow 0$ for the Scalar WGC particle implies a vanishing cut-off $\Lambda \rightarrow 0$ \cite{gpv}. Further, if we accept that there are no additional constraints of relevance on the 5-dimensional theory or the parameter $M_H$, then the Scalar WGC particle must be a Kaluza-Klein mode rather than the particle with the largest charge-to-mass ratio. This tells us that we can not utilise the Scalar WGC to constrain its mass to be below the UV cut-off scale of the theory.

It is interesting to note that both the Scalar WGC $\mu M_p \geq m$ and the original WGC $g M_p \geq m$ can be unified in a sense if we modify (\ref{GWGC}) to 
\be
\left|g^2 -\mu^2 \right| M_p^2 \geq m^2 \;.
\ee
This would be a natural generalisation of (\ref{GWGC}). It would also explain a slight puzzle which is that if we send $M_p \rightarrow \infty$ in (\ref{GWGC}) we seem to still have information left about the sign of $g^2 - \mu^2$.

\section{Quantum Corrections and Naturalness}
\label{sec:QC}

So far we have primarily discussed the evidence for UV/IR mixing in the WGC. In this section we focus on quantum corrections and the implications of the WGC for naturalness. There are two key questions in thinking about this. The first is to do with the fact that (\ref{GWGC}) is taken to apply in the presence of massless scalar fields, but at the quantum level we expect these scalars to gain a mass. This can lead to corrections to (\ref{GWGC}) which need to be quantified. The second question is how to obtain a hierarchy between $\beta$ and $g$ that is protected from quantum corrections. Both of these are model-dependent questions, but there are some general results that we discuss in this section.

\subsection{Quantum Corrections to $m_{\phi}$}

Consider first the issue of the mass of the scalar force mediators. For ease of notation let us consider a single scalar (or pseudo-scalar) mediator $\phi$ with mass $m_{\phi}$. The WGC bound (\ref{GWGC}) applies precisely only for $m_{\phi}=0$. More generally, we can expect it to apply approximately as long as $m_{\phi} \ll m$. This is the regime where the scalar $\phi$ acts as a long range classical force on the WGC particle. One way to ensure a mass separation at the quantum level is to impose supersymmetry. This protects the scalar mass $m_{\phi}$ from perturbative quantum corrections, and as long as non-perturbative effects are small, this is sufficient. Of course, the problem with this is that in protecting $m_{\phi}$ we are also protecting $m$. This makes the WGC bound on $m$ less interesting, but it remains non-trivial. For example, it could be used to explain the smallness of technically natural parameters, such as a supersymmetric mass term. 

To gain insights into naturalness we must consider a non-supersymmetric setting. In this case a massless, or at least arbitrarily light, scalar field at the quantum level is difficult to implement but not impossible. For example, axions can have arbitrarily low masses. More generally, if it couples only very weakly to other sectors its mass could be protected. Also the scalar, or pseudo-scalar, does not have to be fundamental, it could be a type of pion. With this in mind we should recall that, as discussed in section \ref{sec:gwgc}, the pseudo-scalar interaction required is a long range spin-independent one, which means that it relies on some breaking of parity. 

The mass $m_{\phi}$ is therefore model-dependent, and while it may or may not be difficult to protect it, this is a question of how it couples to other fields. However, there is a universal minimum bound on its coupling because in order to apply (\ref{GWGC}) we require that the scalar field $\phi$ couples to the WGC particle with coupling $\mu$ that is not arbitrarily small. Therefore, at the quantum level, we know that its mass at least receives corrections from the WGC particle so that we expect it to be driven to at least $m_{\phi} \sim \mu m$. This is the generic expectation though and it may be possible to construct models where it is much lighter. This is not reintroducing the full hierarchy problem because it is a question of protecting $m_{\phi}$ from $m$ rather than from the full UV physics. We can generate a hierarchy $m_{\phi} \ll m$ by taking $\mu \ll 1$. In general, we expect that we require to take $\mu$ and $g$ very small so as to reduce sensitivity to loops. However, we should not take $\mu$ as small as $\mu \sim \beta$ since then we are lowering the UV cut-off scale (\ref{WGCcutoff}) all the way to $m$. In this sense there is a model-building challenge in coupling $\phi$ much more strongly to $h$ than to UV physics.

An important question in a non-supersymmetric context therefore appears to be what are the quantitative corrections to the WGC (\ref{GWGC}) for a finite $m_{\phi}$ but with $m_{\phi} \ll m$. We do not know, but one estimate could be as follows. Since $m_{\phi} \ll m$ we can still approximate the effect of the scalar $\phi$ as a long range classical Yukawa force. The question of the presence of a bound state is now dependent on the separation scale $l$ between the WGC particles. At separation scales $m_{\phi} l \lesssim 1$ the scalar force is still significant, and we can take $l$ as small as $m^{-1}$. At this scale the Yukawa exponential suppression gives a leading correction to the force
\be
\mu^2 \rightarrow \mu^2 - {\cal O}\left( \frac{\mu^2 m_{\phi}}{m}\right) =  \mu^2 - {\cal O}\left(\mu^3\right)  \;. \label{coress} 
\ee
In the last equality we used the expected contribution to the scalar mass from loops of the WGC particle as discussed  above. Such a correction would not modify the bound as long as $\mu^{\frac32} < \beta \ll \mu$. This is a statement on the absence of a bound state at the smallest distance scale at which the force behaves classically. This is a reasonable guess but, since it is not clear what exactly is the constraint from quantum gravity on the presence of bound states, the distance scale at which to apply it remains an even larger uncertainty. However, this is a limitation on our calculation ability rather than a fundamental obstruction to a small $\beta$. In principle, if we knew the corrected $\mu$ we could tune this against $g$ to reach an arbitrarily small $\beta$ unrelated to $\mu$. There is a weaker bound on $\beta$ relative to $\mu$ coming from quantum corrections to which we now turn.

\subsection{Quantum Corrections to $\beta$}

In this subsection we will consider the quantum corrections to $\beta$. We are particularly interested in whether a hierarchy $\beta \ll g$ could be protected against quantum corrections and so could be technically natural. We will not focus on a possible mechanism to generate the hierarchy in the first place. However, there are a some comments worth making regarding some universal features of the effects of UV physics on $\beta$. The first is that since $g$ is the gauge coupling of a $U(1)$, as long as this remains a $U(1)$ and does not complete into a non-Abelian symmetry, adding a heavy state and integrating it out can only decrease the coupling $g$. The second point is that introducing new scalars that couple to the WGC particle has a tree-level effect on $\beta$ because the scalar force $\mu$ is a sum over all contributions. A scalar field can only lead to an attractive force, and so can only increase $\mu$.\footnote{If the new particle is very heavy there is no sense it which it acts as a classical long range force. Nonetheless, it is natural to expect that integrating it out leads to a positive tree-level contribution to $\mu$.} Therefore, there is a sense in which adding new physics only serves to decrease $\beta$. This may be a possible way to reach a hierarchically small $\beta$ to start off with. 

The primary effect we are interested in are the quantum corrections to $\beta$, which depend on the UV completion. We will not study in detail the possibility that $\beta$ could be protected by a symmetry present in the UV theory that then ensures its technical naturalness. It is not clear what type of symmetry could protect $\beta$. It would have to relate gauge fields and scalar fields. Of course, ${\cal N}=2$ supersymmetry is sufficient in this respect. The interesting thing is that we only requires its action on the bosonic fields. Also the setting of a higher dimensional compactification, as in section \ref{sec:gwgc}, offers some protection against quantum corrections because the gauge and scalar fields are related by a higher-dimensional gauge symmetry. For example, deep UV physics above the Kaluza-Klein scale, where the gauge symmetry is restored, does not correct the mass $m_a$. If the 5-dimensional mass $M_H$ or $M_{\Psi}$ could be kept small, then deep UV physics would also not correct a small $\beta$. In any case, however, threshold corrections from the Kaluza-Klein scale are too large. A symmetry that relates gauge fields and scalars on branes in string theory is T-duality. This is interesting in the sense that this duality also ties to UV/IR mixing. 

There are two ways in which the WGC bound on a particle mass could restrict the mass of a scalar field. One is directly, so if the WGC particle is itself a scalar field, then the WGC directly bounds its mass. The other is indirectly. Say the WGC particle is a fermion that gains a mass from Yukawa coupling to a scalar. Then restricting the mass of the fermion WGC particle restricts the vacuum expectation value of the scalar. Through minimising the scalar potential this indirectly bounds its mass. We will consider each possibility in turn.

\subsubsection{A Scalar WGC Particle}

The case of a scalar WGC particle was discussed briefly in the introduction. Consider the model (\ref{toymo}). We will set $m_{\phi}=0$ for simplicity, and consider no additional terms beyond those in (\ref{toymo}) for now. We are interested in the 1-loop corrections to the gauge coupling $g$ and the parameter $\mu$. Consider starting at some UV scale $M$, and running down to an IR scale $m$, and evaluating the change in the IR values of the parameters. For the gauge coupling this reads 
\be
\delta g \sim g^3 \ln \left(\frac{M}{m}\right)\;. \label{chng}
\ee
Therefore, if we consider $g^2 \lesssim \beta$, this variation will not disturb a hierarchy $\beta \ll g$. In fact, $g$ can only ever decrease as long as UV physics does not involves massive vectors coupling to it, which means that the UV corrections can only ever decrease $\beta$ in any case.

Consider now the parameter $\mu$. To understand its corrections we need to consider the 1PI diagrams correcting the cubic coupling. There are two such diagrams, due to the exchange of $\phi$ and the $U(1)$ photon, and neither is divergent due to the fact that they involve one loop momentum integral but three bosonic propagators. Schematically, for $m \ll M$, the correction takes the form
\be
\delta \mu \sim \mu^3 + \mu g^2 \;.
\ee
Again, these do not disturb a hierarchy a long as $\mu^2 \lesssim \beta$. 

These results tell us that coupling the theory (\ref{toymo}) to new physics does not need to spoil a hierarchy $\beta \ll g$. In this sense $\beta$ is a technically natural parameter. It is of course possible to couple in new physics that does spoil the hierarchy at 1-loop. For example, say we introduced a new scalar $S$ with mass $M_S$ that couples to the WGC particle as 
\be
{\cal L} \supset 2 m \mu_S S h^* h \;.
\ee 
Then we would have
\be
\delta \mu \sim \mu \mu_S^2 \left(\frac{m}{M_S}\right)^2 \;.
\ee
We would therefore require $\mu \mu_S \left(\frac{m}{M_S}\right) \lesssim \beta$ to maintain the hierarchy. Since we need $\mu \gg \beta$ this puts a limit on how strongly $S$ can couple to $h$ and on how light $S$ could be. 

Overall, the question of whether a hierarchically small $\beta \ll g$ can be protected against UV physics is model dependent, but there is no serious obstruction to doing so as far as we can see.

\subsubsection{A Fermion WGC Particle}

The case of a fermion WGC particle is similar. Consider a Dirac WGC particle $\psi$ with action
\be
{\cal L} = \frac{M_p^2}{2} R -\frac{1}{4 g^2} F^2 - \overline{\psi} D \psi - \half \left(\partial \phi \right)^2 - m \overline{\psi} \psi - \half m_{\phi}^2 \phi^2 -  \mu \phi \overline{\psi}  \psi + ...  \;. \label{toymofer}
\ee
Again we have $\delta g \sim g^3$ and $\delta \mu \sim \mu^3 + \mu g^2$, and so $\beta$ is protected. The same analysis holds for the case of a pseudo-scalar coupling ${\cal L} \supset \mu \phi \overline{\psi}  \gamma^5 \psi$. 

In this case the bound (\ref{GWGC}) is on a fermion mass. However, it can be turned into a bound on a scalar mass. Consider setting $m=0$ in (\ref{toymofer}), which is technically natural, and instead adding a scalar $h$ with a potential
\be
{\cal L}_h \supset - \half \left(\partial h \right)^2 - \mu_h h \overline{\psi}  \psi - \left( -\half m_h^2 h^2 + \frac14 \lambda h^4 \right) \;.
\ee
Then minimising the potential we obtain the new mass
\be
m = m_h\frac{\mu_h }{\sqrt{\lambda}} \;.
\ee
We therefore obtain a bound on a scalar mass. Adding the scalar $h$ modifies $\mu$ due to a new 1-loop diagram which means that 
\be
\delta \mu \sim \mu \mu_h^2 \left( \frac{m}{m_h} \right)^2 \;.
\ee
We want to maintain $m \sim m_h$, which can be done by choosing $\lambda \sim \mu_h^2$, and so to control this correction to $\beta$ we need to take $\mu \mu_h \lesssim \beta$.

\subsection{Quantum Corrections in the 5-dimensional Model}
\label{sec:QCtm}

The analysis of the dimensional reduction of the 5-dimensional theory studied in sections \ref{sec:gwgc} and \ref{sec:cutoff} was performed at the classical level.  At the quantum level there are significant changes which exemplify the general problems discussed in this section. The problems begin already with the 5-dimensional theory since it is non-renormalisable, the coupling $g_5$ should become strong and the mass $M_H$ is not protected from the UV scale. Even, if we take the 5-dimensional theory with small $g_5$ and $M_H$ as a given starting point, then the 4-dimensional scalars $a$ and $\varphi$ are expected to gain a mass near the Kaluza-Klein scale from loop effects. Also they would have a potential, in the case of $a$ this would be periodic, which would fix dynamically their values, rather than having them as the free parameters utilised in section \ref{sec:cutoff}. This would imply that reaching a small $\beta$ dynamically is difficult. Further, $\beta$ would receive quantum corrections that would be on top of this dynamical problem. 

The point of sections \ref{sec:gwgc} and \ref{sec:cutoff} was not to argue for a mechanism that can address naturalness questions, but rather to provide some evidence for UV/IR mixing in the WGC. The question we are interesting in is therefore if the physical conclusions reached through the classical analysis could still hold at the quantum level. One way to ensure this is to employ supersymmetry. If we consider ${\cal N}=1$ supersymmetry in the 5-dimensional theory then the quantum problems discussed above are all addressed. Consider the 5-dimensional action (\ref{5daction}). In the case of $M_H=0$ it forms the bosonic sector of an ${\cal N}=1$ supersymmetric theory. Therefore we could complete it to a supersymmetric theory and the analysis of sections \ref{sec:gwgc} and \ref{sec:cutoff}  would remain unchanged. 

In such a supersymmetric setting the ${\cal N}=1$ 5-dimensional supersymmetry reduces on a circle to ${\cal N}=2$ 4-dimensional supersymmetry. This explains the vanishing of (\ref{4dbo}) since $h$ becomes BPS. It is still the case that the WGC bounds the mass of the particle below the UV scale of the theory, although this only amounts to the statement that it should be BPS.

Now consider still a supersymmetric 5-dimensional action apart from turning on $M_H \neq 0$. This is a non-supersymmetric mass term and so breaks supersymmetry completely.  However, since it is the only source of supersymmetry breaking it forms a control parameter for its effects. This setup therefore allows us to perturb away from ${\cal N}=2$, and we see this since (\ref{4dbo}) no longer vanishes. It is reassuring that under this non-supersymmetric perturbation we recover the 5-dimensional WGC. From the 5-dimensional perspective a small $M_H$ is fine since it is technically natural. The 4-dimensional scalars will still obtain a potential, and $\beta$ quantum corrections. However, we are not after a mechanism that dynamically fixes a small $\beta$. We therefore assume that as long as the masses $m_a$ and $m_{\varphi}$ are sufficiently light we could imagine displacing the fields to the point of small $\beta$ by hand. So that it still makes sense to ask how the theory would behave at small $\beta$. With this assumption, we are left with the question of the scalar masses. We require a hierarchy $m_a  \ll m$ and $m_{\varphi} \ll m$ to be able to utilise the WGC. Such a hierarchy can be induced by the general mechanism described in this section. Since their masses are induced by $M_H$ as the source of supersymmetry breaking, they are of order $\mu_a M_H$ and $\mu_{\varphi} M_H$. While $m$ is at least of order $M_H$. We therefore reach the situation described in (\ref{coress}). If indeed the corrections to the WGC are as proposed in (\ref{coress}), then for sufficiently small couplings these are sufficiently small to trust the results. 

We therefore conclude that, under some reasonable assumptions, the classical analysis of sections \ref{sec:gwgc} and \ref{sec:cutoff} is expected to capture the correct physics as long as $\mu_a$ and $\mu_{\varphi}$ are small. 

\section{Discussion}
\label{sec:summary}

In this note we proposed that in the presence of scalar fields the general version of the WGC (\ref{exmod}) can bound the mass of the WGC particle far below the UV cut-off scale of the effective theory. We supported this by presenting new evidence for (\ref{exmod}) and for the claim that taking small $\beta$ does not necessarily lower the cut-off scale of the effective theory. Such a bound on an IR mass from UV quantum gravity physics is interesting in that it manifests a form of UV/IR mixing. Of course, our results depend on the validity of (\ref{exmod}) and while we presented some new arguments for it, extending the work in \cite{Palti:2017elp}, it is crucial to build up more evidence for it. 

A primary motivation for the interest in such a bound is that it could have potential relevance to the question of the naturalness of a scalar mass. At a general level this could be tied to UV/IR mixing in quantum gravity and so any insights into this are useful. In this respect, it was interesting to see that in the toy model 5-dimensional reduction on a circle example, the UV/IR mixing was tied to the interaction of the WGC with a Wilson line in the extra dimensions which is a non-local operator.  

A direct application of our current understanding of the bound from the WGC to the naturalness problem of a scalar mass requires addressing two key questions. The first is how to protect the mass of the force mediator scalar $m_{\phi}$. The second is how to protect a hierarchy $\beta \ll g$. We presented arguments that while this is a model-dependent question, there is no obstruction in principle to doing so in the sense that the ingredients that go into the bound, as in the toy model (\ref{toymo}), do not in themselves obstruct a small $m_{\phi}$ and $\beta$ at the quantum level. Having stated this, there is a model-building challenge in that $\phi$ must couple much more strongly to $h$ than to UV physics. Also, we only considered if a hierarchy $\beta \ll g$ is technically natural, and did not study in detail how such a hierarchy could be induced in the first place. 

One of the largest uncertainties in our analysis is the magnitude of the correction to (\ref{exmod}) for finite but small $m_{\phi}$. We presented an estimate of this in (\ref{coress}), but further work on understanding this effect is crucial. Our estimate, combined with a natural expectation for the minimal value of $m_{\phi}$, led to a bound on how large $g$ could be of $g \lesssim \beta^{\frac{2}{3}}$. This implies that the $U(1)$ and scalar mediators must be very weakly coupled. It also implies a relatively low UV scale $\Lambda_{UV} \lesssim \beta^{\frac{2}{3}} M_p$. However, these limitations were really due to our lack of knowledge of the precise effect of a finite $m_{\phi}$. In principle, $\beta$ could be tuned accounting for this correction arbitrarily far below $g$. 

Another relation between $g$, $\mu$ and $\beta$ comes from controlling quantum corrections which implies $g^2\sim\mu^2 \lesssim \beta$. This relation implies the less strong bound $\Lambda_{UV} \lesssim \sqrt{\beta} M_p$.

We end with a few brief comments about the mechanism discussed in this note and the actual hierarchy problem as applies to the Higgs in the Standard Model. Since the WGC bound relies on coupling to a $U(1)$ and light scalars, which should be in some hidden sector, it is easier to imagine an indirect bound on the Higgs mass. So that the WGC particle obtains some of its mass from the Higgs vacuum expectation value and thereby restricts it. An attractive feature of the bound on the mass from the WGC is that it involves only IR quantities that are potentially measurable by experiments. So say that some particle had a light mass due to the WGC bound, it would imply a sharp prediction that its couplings to massless scalar fields and gauge fields would have to be equal up to an accuracy of the ratio of its mass to the UV scale. If the scalar fields are not massless but still very light, then there would be small corrections to this equality. One would expect a similar such striking relation between gauge and scalar couplings for some particle, not necessarily the Higgs, possibly dark matter, if this mechanism plays a role in the hierarchy problem of the Standard Model. 

If the scalar force mediators have a significant mass then the required equality of the couplings can be significantly corrected. More generally, the WGC argument is based on a classical long range force picture, but it should be better thought of as an argument for a more general microscopic property of quantum gravity that manifests UV/IR mixing and most likely is present even when the long range classical force picture breaks down. In this more general sense, one could imagine even a very massive scalar mediator leading to the bound on the WGC particle mass. In particular, it could be applied to the Standard Model matter fields with the Higgs as the scalar force mediator. While the mass of the Higgs implies there is no classical long range force picture, the more general mechanism could still manifest as a more complicated relation between the gauge and scalar couplings. A quantitative understanding of this would require understanding the effects of a large mass for the scalar force mediators in (\ref{GWGC}). Such a formulation could then be applied directly to the Standard Model Yukawa and gauge couplings, with say the electron as the WGC particle. If the hierarchy problem is then tied to the WGC, as applied to the Standard Model in this way, then the quantitative formulation would predict a specific relation to high precision between the gauge and Yukawa couplings that could be experimentally tested.

In this note we proposed a new way to think about separation of scales and UV/IR mixing. While we focused on a possible application to the gauge hierarchy problem, it would be very interesting to establish if it also has implications for the cosmological constant problem. 

We have discussed how quantum gravity may ensure the absence of towers of bound states that are protected from decay by their charge. There is another way that a tower of bound states can be protected which is by the particle being the lightest in the theory. In the Standard Model these are neutrinos and therefore neutrinos could form a tower of stable bound states. At long distances the only force that neutrinos feel is gravity. Therefore, in empty space they would indeed form bound states. However, this is not the case in the presence of a cosmological constant. In the weak gravity regime the cosmological constant can be modelled as a repulsive linear force (see for example \cite{dark}) so that the total gravitational force acting on two neutrinos is
\be
F_{\mathrm{Gravity}} = m_{\nu}\left(- \frac{m_{\nu}}{r^2} + \frac{\Lambda r}{3} \right)\;,
\ee
where $m_{\nu}$ denotes the lightest neutrino mass.
We can apply this approximately up to the scale of the neutrino mass $r \sim m_{\nu}^{-1}$. Therefore, for neutrinos to not form stable bound states we require
\be
\Lambda > m_{\nu}^{4} \;.
\ee
This bound can be viewed either as a bound on how small the cosmological constant could be or as a bound on how heavy neutrinos could be. In the latter case, this can be translated to a bound on the scale of electroweak symmetry breaking, and therefore the mass of the Higgs. We therefore find the striking result that the mass of the Higgs could not be much higher else, for the given value of the cosmological constant, neutrinos would form a tower of stable bound states. This could lead to a species problem and in that sense be inconsistent with quantum gravity. It would be interesting to explore a possible connection to the results of \cite{Ibanez:2017kvh,Hamada:2017yji,Ibanez:2017oqr}.

\vspace{0.5cm}
{\bf Acknowledgements:} We would like to thank Gia Dvali, Inaki Garcia-Etxebarria, Arthur Hebecker, Luis Ibanez, Daniel Junghans and Irene Valenzuela for useful discussions. The work of DL is supported by the ERC Advanced Grant ``Strings and Gravity" (Grant
No. 320040).



\end{document}